\pdfoutput = 1
\documentclass[
 aps,
 prl,
 reprint,
 nofootinbib,
 nobibnotes,
 superscriptaddress,
 preprintnumbers,
 groupedaddress
]{revtex4-1}

\usepackage[utf8]{inputenc}
\usepackage[T1]{fontenc}
\usepackage{lmodern}
\usepackage{textcomp}
\usepackage{microtype}

\usepackage[abbreviations, british]{foreign}
\newcommand{\no}[1]{\textnumero~#1}
\newcommand{\software}[2][\hspace{-\fontdimen2\font}]{\texttt{#2~#1}}

\usepackage{mathtools}
\usepackage{mathrsfs}
\usepackage{fixmath}
\usepackage{alphabeta}
\usepackage{bm}
\usepackage{physics}
\usepackage{units}
\usepackage{slashed}
\usepackage{isotope}
\newcommand{\inv}[2][1]{#2\ensuremath{^{-#1}}}

\usepackage{ragged2e}
\usepackage[inline]{enumitem}
\setlist{noitemsep}
\newlist{inlinelist}{enumerate*}{1}
\setlist*[inlinelist,1]{itemjoin={,\ }, itemjoin*={, and\ }, after=.}

\usepackage[subrefformat = parens]{subcaption}

\DeclareCaptionFormat{revtex}{#1#2\justifying{#3}}
\usepackage{booktabs}
\usepackage{dcolumn}
\usepackage{graphicx}
\graphicspath{{./figures/}}
\providecommand{\tikzsetnextfilename}[1]{}
\newcommand{\graphics}[2][1]{\tikzsetnextfilename{#2}\centering\includegraphics[width=#1\linewidth]{#2}}

\usepackage[hidelinks, pdfencoding = auto]{hyperref}
\usepackage[noabbrev, capitalize]{cleveref}
\crefname{enumi}{point}{points}
\makeatletter\DeclareRobustCommand{\labelcrefrange}[2]{\@crefrangenostar{labelcref}{#1}{#2}}\makeatother

\let\oldcite\cite\renewcommand\cite{\unskip~\oldcite}
\let\oldeqref\eqref\renewcommand\eqref{\unskip~\oldeqref}
\let\oldparagraph\paragraph\renewcommand{\paragraph}[1]{\oldparagraph{#1:}}

\begin{document}

\preprint{CP3-18-60}

\title{Searching for New Long Lived Particles in Heavy Ion Collisions at the LHC}

\author{Marco Drewes}
\email{marco.drewes@uclouvain.be}
\author{Andrea Giammanco}
\email{andrea.giammanco@uclouvain.be}
\author{Jan Hajer}
\email{jan.hajer@uclouvain.be}
\author{Michele Lucente}
\email{michele.lucente@uclouvain.be}
\author{Olivier Mattelaer}
\email{olivier.mattelaer@uclouvain.be}

\affiliation{Centre for Cosmology, Particle Physics and Phenomenology, Université catholique de Louvain, Louvain-la-Neuve B-1348, Belgium}

\begin{abstract}
We show that heavy ion collisions at the LHC provide a promising environment to search for new long lived particles in well-motivated New Physics scenarios.
One advantage lies in the possibility to operate the main detectors with looser triggers, which can increase the number of observable events by orders of magnitude if the long lived particles are produced with low transverse momentum.
In addition, the absence of pileup in heavy ion collisions can avoid systematic nuisances that will be present in future proton runs, such as the problem of vertex mis-identification.
Finally, there are new production mechanisms that are absent or inefficient in proton collisions.
We show that the looser triggers alone can make searches in heavy ion data competitive with proton data for the specific example of heavy neutrinos in the Neutrino Minimal Standard Model, produced in the decay of $B$ mesons.
Our results suggest that collisions of ions lighter than lead, which are currently under discussion in the heavy ion community, are well-motivated from the viewpoint of searches for New Physics.
\end{abstract}

\maketitle

\paragraph{Introduction}

The Large Hadron Collider (LHC) at CERN is currently the world's most powerful particle collider.
It was built for three main reasons,
\begin{inlinelist}[label = \alph*)]
\item to unveil the origin of electroweak symmetry breaking (EWSB)
\item to search for New Physics beyond the Standard Model (SM)
\item to study the properties of the quark-gluon plasma (QGP)
\end{inlinelist}
The first goal has been achieved with the discovery of a scalar boson \cite{Aad:2012tfa, Chatrchyan:2012xdj}, commonly referred to as ``Higgs boson'', which (given the properties that have been measured so far) strongly suggests that the Brout-Englert-Higgs mechanism \cite{Englert:1964et, Guralnik:1964eu, Higgs:1964pj} is responsible for the spontaneous EWSB as predicted in the SM.
Heavy ion collisions at the LHC have considerably helped to understand the quark-gluon plasma at high collision energies \cite{Roland:2014jsa}.
However, no new elementary particle other than the Higgs boson has been found to date, neither any other deviation from the SM.
In this Letter, we point out that the heavy ion collisions, originally motivated to study the QGP, can also be exploited to extend the search for new particles to parameter regions that are inaccessible with proton collisions.

The motivations for the existence of new elementary particles are manifold.
Apart from theoretical questions such as the \emph{hierarchy problem} \cite{tHooft:1979rat},
the \emph{strong CP problem} \cite{Belavin:1975fg, tHooft:1976rip, Jackiw:1976pf},
and the \emph{flavor puzzle} \cite{Weinberg:1977hb} in the SM, there are a number of unexplained observational facts.
These include \emph{the origin of neutrino masses}, the observed \emph{Dark Matter} (DM), and the \emph{baryon asymmetry of the universe} (BAU).
Many theories that aim to solve these problems predict the existence of new particles.
From an experimental viewpoint, there could be two explanations why no new particles have been found to date:
Either they are too heavy to be produced in collisions at the LHC at all, or they are too feebly coupled to be produced in significant numbers.
While the former possibility motivates the construction of more powerful colliders (``energy frontier''), the latter option provides a strong incentive to increase the number of collisions (``intensity frontier'') and/or reduce the backgrounds in experiments.
In this Letter, we present a novel approach to explore the second possibility to search for new \emph{long-lived particles} (LLPs).

\paragraph{Long-lived Particles}

Particles that exist for long enough to travel macroscopic (measurable) distances in the LHC experiments appear in many models of physics beyond the SM \cite{Alekhin:2015byh, Curtin:2018mvb, Beacham:2019nyx, Alimena:2019zri}.
This leads to striking \emph{displaced signatures}.
LLPs usually owe their longevity to feeble effective couplings to known particles, often in combination with a comparably low mass.
This is particularly well-motivated in theories with dark or hidden sectors that communicate with the SM only via so-called ``portal'' interactions.
These sectors may include one or several DM candidates, as well as the ingredients to explain the generation of neutrino masses and/or the BAU.
Recently, several proposals have been made to improve the sensitivity of the LHC to very large displacements, including the addition of extra detectors along the beamline or at the surface \cite{Chou:2016lxi, Gligorov:2017nwh, Kling:2018wct, Curtin:2018mvb, Alpigiani:2018fgd}.
In this work, we pursue another direction and explore the idea that the discovery potential of the existing detectors can be improved by searches in heavy ion collisions.

\paragraph{Heavy ion collisions at the LHC}

Heavy ion collisions offer a number of unique advantages that can be utilized for searches at the intensity frontier.
\begin{enumerate}[label = \roman*)]

\item \label{it:interactions} With respect to protons, a larger number of parton level interactions per collision can be achieved for heavy ions, as each nucleus contains $A$ nucleons, providing an enhancement of up to four orders of magnitude in hard-scattering cross sections with respect to $pp$ collisions at the same center-of-mass energy per nucleon ($\approx A^2$, with $A = 208$ for the lead isotope accelerated in LHC beams, \isotope[208][82]{Pb}).

\begin{figure}
\graphics{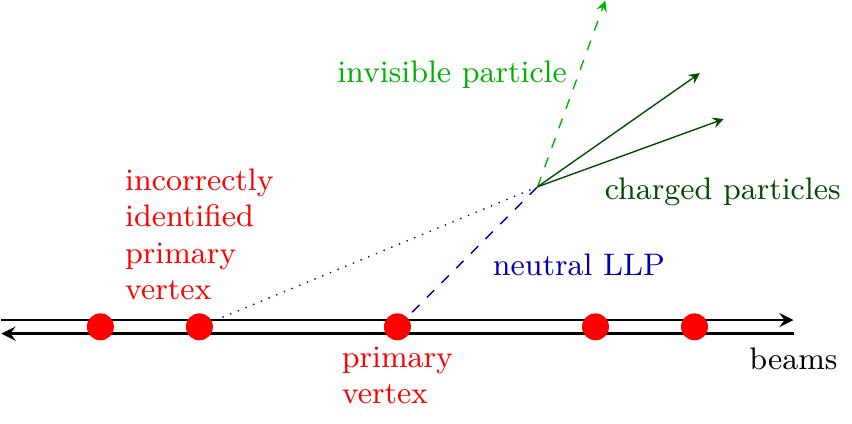}
\caption{
Example of a signature that is difficult to search for in high pileup $pp$ collisions.
Heavy ion collisions can provide a cleaner environment.
} \label{fig:misidentification}
\end{figure}

\item \label{it:pileup} The probability of mis-identifying the primary vertex is practically negligible for heavy ion collisions.
This is in contrast to the pileup that one has to face when colliding high intensity proton beams, which leads to tracks that originate from different points in the same bunch crossing and creates a considerable background for displaced signatures.
Hence, heavy ion collisions provide a much cleaner environment to search for signatures stemming from the decay of LLPs, \cf \cref{fig:misidentification}.

\item \label{it:trigger} The lower instantaneous luminosity can enable ATLAS and CMS to significantly lower their trigger thresholds, in particular for clean analysis objects such as muons.
This, \eg, allows to search for signatures with comparably low transverse momentum $p_T$, which is particularly interesting in scenarios involving light LLPs \cite{Sirunyan:2017oug}.

\item \label{it:fields}
Heavy ion collisions can offer entirely new production mechanisms that are absent in proton collisions.
We do not exploit these here, details and references can \eg be found in \cite{Bruce:2018yzs}.
\end{enumerate}
On the other hand, there are also a number of drawbacks in comparison to proton collisions.
\begin{enumerate}[label = {\arabic*)}]

\item The collision energy per nucleon is smaller (\eg $\sqrt{s_\text{NN}} = \unit[5.52]{TeV}$ for Pb beams), leading to reduced production cross sections with respect to $pp$ collisions at \unit[14]{TeV}.
This effect depends on particle mass (weaker for lighter particles) and colliding partons (stronger for $gg$, weaker for $q\bar q$).

\item Heavy ion collisions produce a huge number of tracks near the interaction point.
While this makes it very difficult to search for prompt signatures, feebly interacting LLPs can leave the luminous region before decaying.
Their decay can produce displaced tracks at macroscopic distances from the interaction point that can easily be distinguished from the tracks that originate from the collision itself.
The comparison of $b$-tagging performance in $t\bar t$ studies in $pp$, $p$Pb, and PbPb collisions \cite{SiRunyan:2017xku, CMS-PAS-HIN-19-001} indicates that an acceptable purity versus efficiency can be achieved even in conditions of very large track multiplicity in heavy ion data.
Moreover, due to the high pileup during Run~4 in $pp$ collisions, the track multiplicity in ion collisions is expected to be comparable for $pp$ and PbPb and even smaller for lighter ion beams \cite{Sirunyan:2019cgy}.
In both $pp$ and heavy ion collisions, secondary nuclear interactions in the detector material constitute a source of background for this kind of signal.

\item The instantaneous luminosity in heavy ion runs is considerably lower than in $pp$ collisions.
During the latest PbPb Run, in late 2018, the LHC delivered \unit[1.8]{\inv{nb}} of collisions to the ATLAS and CMS detectors, while \unit[10]{\inv{nb}} are expected to be accumulated during the high-luminosity phase of the accelerator (HL-LHC) \cite{Citron:2018lsq}.
This poses the strongest limitation to displaced vertex searches.
We discuss it in some detail in the following.

\item Heavy ion Runs at the LHC are relatively short (not more than one month is allocated in the yearly schedule).
To remain independent of possible changes in the planning, in the following we compare the sensitivity per running time, given a realistic instantaneous luminosity.
\end{enumerate}
In this Letter, we show that there are well-motivated scenarios in which searches in heavy ion collisions can be competitive with proton collisions.
The effect of \cref{it:pileup,it:fields} are model dependent and require a detailed quantitative analysis.
This goes beyond the scope of the present Letter, whose main purpose is to point out the potential of heavy ion collisions.
Therefore, we take a conservative approach and entirely focus on \cref{it:interactions,it:trigger} for illustrative purposes.

\paragraph{Average instantaneous luminosity}

The upper limit on the achievable instantaneous luminosity depends on the mass number $A$ and charge $Z$ of the concerned nuclei in a complicated manner and is currently under investigation.
For the purpose of the present Letter we compute the optimized luminosity based on the numbers given in \cite{Citron:2018lsq}.
We present details of this computation in \cite{Drewes:2019vjy}.
In the following we briefly summarize how we used these data.
The instantaneous luminosity at an interaction point behaves as $\mathscr L \propto n_b N_b^2$ \cite{Benedikt:2015mpa}, where $n_b$ is the number of bunches per beam and $N_b$ is the number of nucleons per bunch.
The initial bunch intensity roughly follows $N_b \left(\isotope[A][Z]{N}\right) = N_b \left(\isotope[208][82]{Pb}\right) \left(\frac{Z}{82}\right)^{-p}$ where the exponent characterizes the number of nuclei per bunch.
For a given isotope, it is limited by the heavy-ion injector chain, the bunch charges and intra-beam scatterings.
Simple estimates based on fixed target studies with Ar beams suggest that $p \lesssim 1.9$ is realistic \cite{Citron:2018lsq}.
Therefore, $\mathscr L$ becomes smaller if the ions are heavier, suggesting that ions lighter than lead might be optimal for searches of LLP.

\paragraph{An example: Heavy Neutrinos}

Right handed neutrinos $\nu_R$ appear in many extensions of the SM and could, depending on their masses, explain several open puzzles in cosmology and particle physics, \cf \eg \cite{Drewes:2013gca}.
In order to describe collider signatures with displaced vertices it suffices to consider a single neutrino generation $\nu_R$ with Majorana mass $M$.
The most general renormalizable extension of the SM including $\nu_R$ resembles the type-I seesaw The Lagrangian \cite{Minkowski:1977sc, GellMann:1980vs, Mohapatra:1979ia, Yanagida:1980xy, Schechter:1980gr, Schechter:1981cv}.
The Lagrangian that describes the interaction of the heavy neutrino mass eigenstate $N \simeq \nu_R + \theta_a \nu_{La}^c + \text{c.c.}$ with the SM reads
\begin{multline}
 \mathcal L
 \supset
- \frac{m_W}{v} \overline N \theta^*_a \gamma^\mu \ell_{L a} W^+_\mu
- \frac{m_Z}{\sqrt 2 v} \overline N \theta^*_a \gamma^\mu \nu_{L a} Z_\mu \\
- \frac{M}{v} \theta_a h \overline{\nu_L}_\alpha N
+ \text{h.c.}
\ ,\label{eq:weak intraction}
\end{multline}
with $\theta_a = \flatfrac{v F_a}{M}$, where $v \simeq \unit[174]{GeV}$ is the Higgs field expectation value in vacuum and $h$ is the physical Higgs field after spontaneous breaking of the electroweak symmetry.
$F_a$ is a Yukawa coupling with flavor index $a$, $\ell_L$, and $\nu_L$ represent the charged leptons and neutrinos, $m_Z$ and $m_W$ are the masses of the weak gauge bosons $W^\mu$ and $Z^\mu$, respectively.
Lagrangian \eqref{eq:weak intraction} effectively describes the phenomenology of the Neutrino Minimal Standard Model ($\nu$MSM) \cite{Asaka:2005an, Asaka:2005pn}, a minimal extension of the SM that can simultaneously explain the light neutrino masses, the BAU, and the DM \cite{Canetti:2012vf, Canetti:2012kh}, \cf \cite{Boyarsky:2009ix,Drewes:2013gca} for reviews.

\begin{figure}
\begin{subfigure}{.49\linewidth}
\centering
\graphics{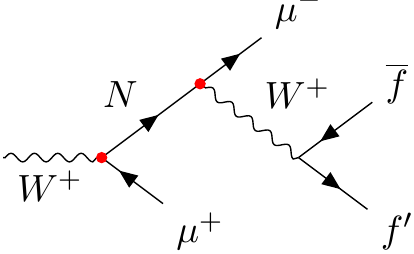}
\caption{
Electroweak production
} \label{fig:boson feynman}
\end{subfigure}
\hfill
\begin{subfigure}{.49\linewidth}
\centering
\graphics{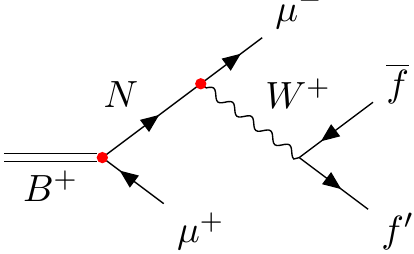}
\caption{
$B$-meson production
} \label{fig:meson feynman}
\end{subfigure}
\caption{
Production and decay of heavy neutrino mass eigenstates $N$.
The tiny couplings inducing the displaced signature are indicated by red vertices.
} \label{fig:feynman}
\end{figure}

For $M < m_W$ the heavy neutrinos can be long-lived enough to produce displaced vertex signals.
The dominant production channel for $M > \unit[5]{GeV}$ is the decay of real $Z$ ($W$) bosons, in which the heavy neutrino $N$ is produced along with a neutrino $\nu_a$ (charged lepton $\ell_a$), while for $M < \unit[5]{GeV}$ the production in $b$ hadron decays dominates, \cf \cref{fig:feynman}.
The production cross section for heavy neutrinos can be estimated as $\sim L \sigma_\nu U_a^2$, where $L$ is the integrated luminosity, $\sigma_\nu$ is the production cross section for light neutrinos, and $U_a^2 = \abs{\theta_a}^2$.
It is given by $\sigma_\nu = \sigma_W / 3$ in $W$ decays and $\sigma_\nu = \sigma_B / 9$ in $B$ decays, where $\sigma_W$ and $\sigma_B$ are the $W$ and $B$ production cross sections.
The $N$ then decay (semi-)leptonically \cite{Gorbunov:2007ak, Atre:2009rg, Canetti:2012kh, Bondarenko:2018ptm}.

For simplicity, we assume that $N$ exclusively mixes with the second generation ($a = \mu$).
The number of displaced vertex events that can be seen in a spherical detector is given by
\begin{equation}
N_\text{obs} \simeq L \sigma_\nu U_\mu^2 \left(e^{\nicefrac{-l_0}{\lambda}} - e^{\nicefrac{-l_1}{\lambda}} \right) f_\text{cut}
\label{eq:number W events}
\end{equation}
where $l_1$ is the length of the effective detector volume, $l_0$ the minimal displacement that is required by the trigger, $\lambda = \flatfrac{\beta \gamma}{\Gamma_N}$ is the particle decay length, with $\Gamma_N$ the heavy neutrino decay rate, $\beta$ the heavy neutrino velocity, $\beta \gamma = \flatfrac{\abs{\bm p}}{M}$ are Lorentz factors, and $f_\text{cut} \in [0, 1]$ is an overall efficiency factor related to triggers, detector geometry, and selection.

We focus on the $N$ production in $b$ hadron decays because here one can expect a considerable improvement, as most $b$ hadrons are produced with low $p_T$.
We estimate the number of events with the simple detector model \eqref{eq:number W events} with $\Gamma_N \simeq 11.9 \cdot \nicefrac{G_F^2}{96\pi^3} U_\mu^2 M^5$ \cite{Gorbunov:2007ak, Canetti:2012kh}.
We fitted $l_0 = \unit[2]{cm}$, $l = \unit[20]{cm}$, and $f_\text{cut} = 0.1$ by comparing to a simulation of
displaced decays of $N$ from $W$ bosons in heavy ion collisions, as explained in detail in \cite{Drewes:2019vjy}.
The momentum average is performed using distribution functions which we generated by simulating the process $pp \to W$ with up to two jets using \software[2.6.4]{MadGraph5\_aMC@NLO} \cite{Alwall:2014hca} with subsequent hadronization and matching with soft jets via \software[8.2]{Pythia} \cite{Sjostrand:2014zea}.
The simplified model reproduces the number of events in $pp$, ArAr, and PbPb collisions in the simulation within a factor of order one in most of the parameter space.
We then use \eqref{eq:number W events} with these values to predict $N_\text{obs}$ for the production in $B$ meson decays with $\sigma_\nu = \sigma_B / 9$.
We compute the differential $B$ meson production cross section at next-to leading order (NLO) with next-to leading logarithmic terms (NLL) in the \software{FONLL} framework \cite{Cacciari:2015fta}, using a value $f(b \to B^+) = 0.403$ for the $b$-quark fragmentation fraction \cite{Cacciari:2012ny}, a pseudorapidity of $\abs{\eta} < 4$, the \software[6.6]{CTEQ} NLO parton distribution functions \cite{Nadolsky:2008zw}, and we employ the nuclear modification factors measured in \cite{Sirunyan:2017oug}.
We can use the $B$ meson $p_T$ distribution as a very conservative estimate for that of the primary muons \cite{Drewes:2019vjy} and employ a cut of $p_T > \unit[25]{GeV}$ in $pp$ collisions and of $p_T > \unit[3]{GeV}$ in heavy ion collisions, the latter being the minimum transverse momentum that allows a charged particle to cross the muon chambers of the CMS experiment.

\begin{figure}
\graphics{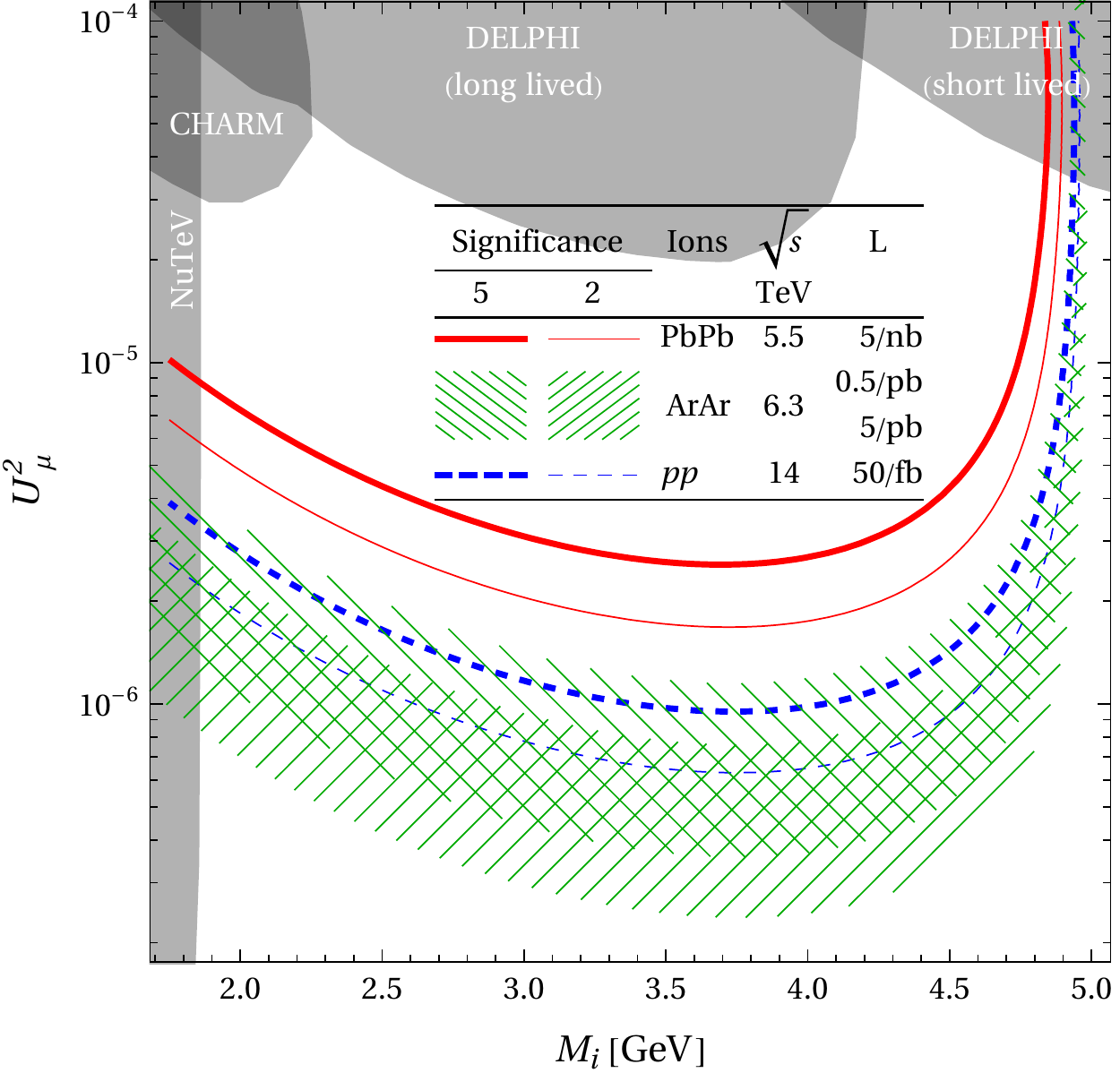}
\caption{
CMS reach in the heavy neutrino mass $M$ and coupling $U_\mu^2$ plane for heavy neutrinos produced in $b$ hadron decays in collisions of $pp$ at \unit[14]{TeV} (dashed blue), ArAr at \unit[6.3]{TeV} (green hashed areas reflecting the uncertainty in the achievable beam intensity \cite{Citron:2018lsq, Drewes:2019vjy}), and PbPb at \unit[5.52]{TeV} (solid red).
The values for the integrated luminosity $L$ are given by requiring one month of run time.
The thin and thick curves correspond to 4 and 9 signal events, \ie \unit[2 and 5]{\sigma} standard deviations, respectively.
In the entire white region heavy neutrinos could simultaneously explain the light neutrino masses and the baryon asymmetry of the universe \cite{Abada:2018oly}.
} \label{fig:meson sensitivity}
\end{figure}

For this study, we ignored systematic uncertainties, as the main limitation to sensitivity is assumed to come from the size of the dataset.
However, under real analysis conditions the following sources of uncertainty will need to be estimated
\begin{inlinelist}[label = \alph*)]
\item modelling of the spectrum of $B$ mesons
\item efficiency of the muon trigger, reconstruction, and identification (both for the prompt and for the displaced muon candidate)
\item in the case of $pp$ collisions, the probability of correctly identifying the primary vertex of the hard collision
\item the adverse effect of pileup on the sensitivity in $pp$ collisions
\end{inlinelist}
The last two will work in favor of heavy ion collisions when compared to $pp$ collisions.

\cref{fig:meson sensitivity} shows that searches in PbPb collisions at the HL-LHC could improve the existing constraints by roughly an order of magnitude.
If ions lighter than Pb are collided the gain in sensitivity due to low $p_T$ events can overcompensate the effect of the smaller instantaneous luminosity, leading to a better sensitivity per running time than in proton collisions.
The sensitivity in this regime is larger than what can be achieved with so-called ``parked data'' \cite{CMS-DP-2012-022, CMS-DP-2019-bparking, Drewes:2019vjy}.
Searches in pileup data as advocated in \cite{Nachman:2016nes} may add a significant amount of statistics to the $pp$ dataset \cite{Drewes:2019vjy} but are affected by pileup and other limitations of high-luminosity $pp$ runs.
When comparing to proposed future detectors, ArAr collisions with $L = \unit[5]{\inv{pb}}$ could achieve a higher sensitivity than FASER2 with $L = \unit[3]{\inv{ab}}$ and a comparable sensitivity as CODEX-b with $L = \unit[300]{\inv{fb}}$, while MATHUSLA with $L = \unit[3]{\inv{ab}}$ or the SHiP experiment with $2 \cdot 10^{20}$ would be more sensitive \cite{Beacham:2019nyx}.
However, so far no decision has been made about the construction of SHiP, MATHUSLA and CODEX-b, while for FASER currently only Phase~1 is approved.

\paragraph{Discussion and conclusions}

We studied the potential to search for LLPs via displaced vertex searches in heavy ion collisions.
Heavy ion collisions offer a number of unique advantages compared to $pp$ collisions, including the absence of pileup, a more reliable identification of the primary vertex, the possibility to operate the main detectors with lower triggers, and new production mechanisms.
On the other hand, the average integrated luminosity per unit of running time is considerably lower than in proton collisions.
As a proof of principle, we studied the specific case of heavy neutrinos to show that there are well-motivated scenarios in which heavy ion collisions can probe parameter regions that are difficult to access in proton collisions.
Our analysis is conservative in the sense that it only uses the advantage of looser $p_T$ triggers in heavy ion collisions, leaving a detailed study of the other aspects to future work.
In particular, we did not take the adverse effect of pileup in proton collisions into account, which we expect to be subdominant in the model under consideration.
In models with more complicated event topology and backgrounds we expect that the lack of pileup will be a major advantage of heavy ion collisions with respect to proton collisions at high intensity.

We find that the number of heavy neutrinos produced per unit time is always smaller in heavy ion collisions than in proton collisions.
However, for heavy neutrinos with masses below that of the $B$ mesons the number of events per unit of running time that can actually be observed with realistic selection thresholds can be larger than in proton collisions if ions lighter than Pb are used.
The use of lighter ions is currently explored by the heavy ion community for other reasons \cite{Citron:2018lsq}.
Based on the numbers used here, which are optimistic with respect to beam performance but conservative regarding the event kinematics, this would bring the sensitivity of the scheduled heavy ion collisions in Run~3 in terms of heavy neutrino mixing to the same order of magnitude as that of the $pp$ collisions even if one only exploits only one of the benefits \labelcrefrange{it:pileup}{it:fields}, the possibility to use lower $p_T$ cuts.
Taking advantage of all benefits that heavy ion collision offer can potentially make them more sensitive to LLPs than $pp$ collisions even with the current schedule for the HL-LHC, with further improvement possible if the heavy ion runs were extended, in particular in scenarios where the absence of pileup in heavy ion collisions and the additional production mechanisms play a bigger role than in the conservative model considered here.
Moreover, the QGP generated in heavy ion collisions roughly resembles the primordial QCD plasma in the early universe, allowing to probe the properties of cosmologically motivated LLPs throughout the epoch of hadronization.
For heavy neutrinos this could shed light on the generation of lepton asymmetries \cite{Shaposhnikov:2008pf, Canetti:2012vf, Canetti:2012kh} that lead to a resonant production of Dark Matter \cite{Shi:1998km, Asaka:2006nq}.
For other LLPs, \eg some axion like particles or sexaquarks, this may allow to directly probe the cosmological production mechanism \cite{Bruce:2018yzs}.

In summary, we find that LLP searches in heavy ion collisions can help to explore parameter regions in well-motivated theories beyond the SM that are difficult to access in proton collisions.
Together with a number of previous proposals \cite{Bruce:2018yzs}, these results provide strong motivation to further study the potential to use heavy ion collisions at the LHC to search for New Physics, either by using the data that will be collected in the scheduled runs
or in dedicated runs.

\begin{acknowledgments}

\paragraph{Acknowledgments}

We thank Georgios Konstantinos Krintiras, John Jowett, Fabio Maltoni, Emilien Chapon, Jessica Prisciandaro, Mauro Verzetti, Martino Borsato, Elena Graverini, and Giacomo Bruno for very helpful discussions as well as Albert de Roeck and Steven Lowette for pointing us to relevant material.
This work was partly supported by the F.R.S.-FNRS under the \emph{Excellence of Science} (EOS) project \no{30820817} (be.h).
This project has received funding from the European Union’s Horizon 2020 research and innovation programme under the Marie Skłodowska-Curie grant agreement \no{750627}.

\end{acknowledgments}

\raggedright\bibliography{bibliography}

\end{document}